\documentstyle[twoside,epsf,fleqn,espcrc2]{article}

\def\ai{\'{\i}}
\def\bc{\begin{center}}
\def\ec{\end{center}}
\def\be{\begin{equation}}
\def\ee{\end{equation}}
\def\myappendix{\par
 \setcounter{section}{0}
 \setcounter{subsection}{0}
 \setcounter{equation}{0}
 \setcounter{table}{0}
 \def\appendixname{Appendix}
 \def\appesection{\setcounter{equation}{0}\section}
 \def\@thesection{\Alph{section}}
 \def\thesection{\appendixname\hskip 1.10ex\Alph{section}}
 \def\thesubsection{\@thesection.\arabic{subsection}}
 \def\theequation{\@thesection.\arabic{equation}}
 \def\thetable{\@thesection.\arabic{table}}}

\newcommand{\beq}{\begin{equation}}
\newcommand{\eeq}{\end{equation}}
\newcommand{\beqn}{\begin{eqnarray}}
\newcommand{\eeqn}{\end{eqnarray}}

\def\vdir{v\kern-7.8pt\Big{/}}
\def\pdir{p\kern-7.8pt\Big{/}}

\newcommand{\amb}{\alpha _{s}(\overline{m}_{b})}
\begin{document}
\title{NNLO unquenched calculation of the $b$ quark mass.}

\author{V.~Gim\'enez\address{Dep. de F\ai sica Te\`orica and IFIC, Univ. de Val\`encia,
E-46100, Burjassot, Val\`encia, Spain.}\thanks{Talk presented by V.~Gim\'enez},
L. Giusti\address{Department of Physics, Boston University, Boston, MA 02215 USA},
F. Rapuano\address{Dip. di Fisica, Univ. di Roma ''La Sapienza'' and INFN, 
Sezione di Roma, P.le A. Moro 2, I-00185 Rome, Italy} and G. Martinelli$^{c}$}
\begin{abstract}
By combining the first unquenched lattice computation of 
the B-meson binding energy with the recently calculated two--loop contribution to 
the lattice HQET mass, we determine the 
$\overline{MS}$ b--quark mass, $\overline{m}_{b}(\overline{m}_{b})$ at the 
NNLO. We find $\overline{m}_{b}(\overline{m}_{b}) = 
 (4.26\pm 0.03 \pm 0.04 \pm 0.05)\;\;\mbox{\rm GeV}$.
The inclusion of the two--loop effects is one of the steps necessary 
to extract $\overline{m}_{b}(\overline{m}_{b})$ with a precision better than 
${\cal O}(\Lambda_{QCD})$, which is the uncertainty due to the presence of an IR 
renormalon singularity in the perturbative series of the residual mass.
Our results have been obtained on a sample of $60$ lattices of size 
$24^{3}\times 40$ at $\beta=5.6$, using the unquenched Wilson action with
two degenerate sea quarks. The quark propagators have been computed using the
unquenched links generated by the T$\chi$L Collaboration. 
\end{abstract}
\maketitle
\section{Introduction and Motivation.}
\label{motiv}
Quark masses are fundamental parameters of QCD which cannot be determined by 
theoretical considerations only. Moreover, they are very important for 
phenomenology because enter many theoretical predictions of physical 
quantities like the CKM matrix elements, the CP violation parameter 
$\epsilon'/\epsilon$, the $\Delta I=1/2$ $K\rightarrow \pi\pi$ amplitude, the 
B meson semileptonic decays, the $B-\bar{B}$ mixing \ldots etc.
Quark masses cannot be measured
directly since quarks are confined in the hadrons. Therefore, a short 
distance definition of the quark mass, which is scale and scheme dependent, 
must be adopted.

In this paper, we report on the first unquenched HQET lattice calculation
of the $b$ quark mass. 
The main idea~\cite{ours} is to combine unquenched HQET lattice 
computations of the B-meson propagator with recent NNLO analytical formulae of 
the matching of the continuum $\overline{MS}$ quark mass to the lattice 
one~\cite{MS}. We stress that both the unquenched lattice
simulation and the NNLO matching are necessary ingredients to improve the
accuracy of the results: the former is necessary to control potentially large 
vacuum polarization contributions to the B meson propagator and the latter is 
crucial to cancel the renormalon ambiguities in the matching~\cite{MS}.

\section{How to determine the b-quark mass on the lattice.}
The key idea is to match the $b$-quark propagator in QCD to its lattice HQET
counterpart. Since the lattice HQET is an effective theory,
the relation between these propagators is, to lowest order in $1/m_b$,  
\be\label{eq:matching}
S^{-1}(p,m_{b};\mu)\, =\, C(m_b/\mu,\alpha_s)\, S^{-1}_{L}((v\cdot k)a;\mu)
\ee
where $p=m_{b}\, v+k$ is the external momentum of the $b$-quark, $v$ is its 
velocity, $k$ is the residual momentum with $|k|\ll m_b$, $\mu$ is 
the renormalization point, $a$ is the lattice spacing  and 
$C(m_b/\mu,\alpha_s)$ is the Wilson coefficient. 
Note that, as expansion parameter of the HQET, we have chosen the quark mass
$m_b$. In this way, all
the mass dependence is factorized in the coefficient function $C$. The 
method to fix $C$ is well known: it consists in calculating the $b$-quark 
propagator in QCD and 
in the lattice HQET to a given order in $\alpha_s$, comparing both expressions 
at a fixed $\mu=\mu_0$ and extracting $C(m_b/\mu_0)$. Renormalization group 
can then be used to evolve this function to lower scales. The important point is 
that by rewriting eq.~(\ref{eq:matching}) in terms of the pole mass, 
$m_{pole}$, is easy to find the relation
\be\label{eq:pole}
m_{pole}\, 
=\, 
m_{b}\, +\, \sum ^{\infty }_{n=0}\, \left( \alpha_{s}(a)\right)^{n+1}\, \frac{X_{n}}{a}
\ee
where the last term is the perturbative expansion of the residual mass generated
in the lattice HQET, $\delta m$. The coefficients $X_{n}$ are functions of
 $\ln(m_{b}a)$.  
 
On the other hand, the HQET mass formula, to lowest in $1/m_{b}$,
\be
M_{B}\, -\, {\cal E}\, =\, m_{b}\, +\, {\cal O}(1/m_{b})
\ee
allows us to write the unknown mass $m_b$ in eq.~(\ref{eq:pole}) in terms of the 
physical B-meson mass, $M_{B}$. The so-called binding energy, 
${\cal E}$, is independent of $m_{b}$
\be\label{eq:pole2}
m_{pole} \, =
\, M_{B}\,-\, {\cal E}\, +\, \sum ^{\infty }_{n=0}\, 
\left(\alpha_{s}(a)\right)^{n+1}\, \frac{X_{n}}{a}\;\mbox{\rm .}
\ee
${\cal E}$ is not a physical quantity because it diverges linearly as
$a\rightarrow 0$. Since the pole mass is finite in this limit, this divergence
is cancelled by the last term of eq.~(\ref{eq:pole2}). 
In practice this cancellation is not perfect and hence one cannot take $a$ too
small. As a large $n_{f}$ calculation demostrates (see \cite{beneke}), 
the series in eq.~(\ref{eq:pole2}) has IR renormalon 
singularities, the same as the pole mass. In other words, the coefficients
$X_{n}$ grow as $n!$ as $n\rightarrow \infty$. This behaviour gives rise to 
ambiguities of ${\cal O}(\Lambda_{QCD})$. In order to avoid this
problem, it is 
convenient to use a short distance definition of the
b-quark mass, like the $\overline{MS}$, $\overline{m}_b$, which is free of IR 
renormalons. The relation between the pole and the $\overline{MS}$ mass is known to 
$O(\alpha_s^3)$ \cite{3loops} and has an IR renormalon that cancels the one in (\ref{eq:pole2}).
The cancellation, however, is delicate so that one need to know many 
coefficients $X_{n}$ to keep under control the partially removed renormalon 
ambiguity. $X_{0}$ is obtained easily in terms of a one-loop lattice integral. 
Recently, Martinelli and Sachrajda have performed the calculation 
of the coefficient $X_{1}$~\cite{MS}. For the Wilson action, they are
\begin{eqnarray}
X_{0} &=& 2.1173\nonumber\\
X_{1} &=& (\, 3.707\, -\, 0.255\, n_{f})\, \ln (\overline{m}_{b}a) \nonumber \\
&-&  (\, 1.306\, +\, 0.104\, n_{f})
\end{eqnarray}
Putting all together, we find 
\begin{eqnarray}\label{eq:massa}
\hspace{-0.65cm}&\mbox{}&\overline{m}_{b}(\overline{m}_{b}) = \left( M_{B}\, -\, {\cal E}\, +\, \frac{2.1173}{a}\, \amb\right. \nonumber\\
\hspace{-0.65cm}&\mbox{}& \left. +\, \frac{1}{a}\left[ 3.197\, \ln (\overline{m}_{b}a)\, -\, 1.514\right] \, \amb ^{2}\right) \nonumber\\
\hspace{-0.65cm}&\mbox{}& \times  \left[ 1\, -\, \frac{4}{3}\, \frac{\alpha _{s}(\overline{m_{b}})}{\pi }\, 
-\, 9.58\, \left( \frac{\alpha _{s}(\overline{m_{b}})}{\pi }\right) ^{2}\right]
\end{eqnarray}
where we have taken $n_{f}=2$, the number of sea quarks in our
simulation. Note that the remaining ambiguity in our result is only 
${\cal O}(\Lambda_{QCD}^{2}/\overline{m}_{b})$ which is beyond our precision. 
By computing ${\cal E}$ through lattice simulations of the HQET and 
by using the experimental values of the B-meson masses, 
we can extract $\overline{m}_b$.
\section{Lattice computation of ${\cal E}$.}
As well known, ${\cal E}$ can be extracted by studying the large time 
behaviour of the two--point function of the B meson, $C_{2}$ in the HQET, 
\be\label{eq:larget}
C_{2}(t)\, \longrightarrow \, Z\, \, e^{-{\cal E}\, t} 
\ee
The light quarks are described by the Wilson action.
In order to improve the isolation of the ground state, we use cube and double cube
smeared axial currents as interpolating operators of the B meson~\cite{ours}. 
The actual value of ${\cal E}$ has some dependence on the cube size and 
the smearing type due to contamination from excited states. 
To obtain our best estimate, we compare different methods and account this 
systematic effect in the final error. We use the Standard Method, 
in which we base our results on the best cube, defined as
the one which yields the largest and flattest effective mass plateau. 
In practice, we have few cube sizes (only two in our simulation, $7$ and $9$), 
and hence the best cube is difficult to find. To improve our results,
we also use the Multifit Method which consists in performing a global fit of 
the data for all smearing types and cubes sizes imposing that the binding 
energy be the same for all of them.
In order to reduce the effect of excited states, it is convenient to fit 
the data to a two-state form of $C_{2}$ rather than to eq.~(\ref{eq:larget}).
\section{Analysis and results.}
We performed an unquenched Wilson simulation with two degenerate sea quarks 
at two values of their mass, $k_{sea}=0.1575$ and $k_{sea}=0.1580$, 
at $\beta=5.6$ on a $24^{3}\times 40$ lattice.
The sample is of $60$ configurations at four values of the valence 
light quark masses. We compute the quark propagators and correlation functions
from the gluon configurations generated by the $T\chi L$ Collaboration. 
Since our simulation is unquenched, the procedure to measure
the lattice quantities, is slightly different from the usual quenched case. 
Since there is some confusion in the literature, we want to stress a point
that we consider important in unquenched analyses. We think that the correct
strategy consists in performing an 
independent quenched-like study of all lattice quantities, including the lattice
spacing, for each $k_{sea}$. Only at this point, the quantities expressed in physical 
units can be extrapolated as a function of the sea quark masses. 
The reason is that a change of the sea quark mass  modifies the value of the 
coupling constant, and hence, may induce a rapid variation
of the value of the lattice spacing. Therefore, 
lattice results from different values of $k_{sea}$ are not directly comparable until 
they have been converted to physical units.
We find, from $m_{K^{*}}$,
\be
a^{-1}|_{m_{K^{*}}}\, =\, \{\, 2.51(6), 2.54(6)\, \}\;\;\mbox{\rm GeV}
\ee
and for the binding energy
\begin{eqnarray}
a{\cal E}_{B_{d}} &=& \{\, 0.588(11)(5), 0.606(15)(2)\, \}\nonumber\\
a{\cal E}_{B_{s}} &=& \{\, 0.620(8)(4), 0.632(12)(2)\, \}
\end{eqnarray}
for the two values of $k_{sea}$. More details will be given in~\cite{futuro}.

We have carefully studied different sources of systematic errors in 
eq.~(\ref{eq:massa}). We find that the dependence of 
$\overline{m}_{b}(\overline{m}_{b})$ on the unknown $\Lambda_{QCD}^{n_{f}=2}$ 
is small. Since the quenched value is
$\Lambda_{QCD}^{n_{f}=0}\sim 250$~MeV \cite{lusher}, and the physical one
is expected to be larger, we varied the value of 
$\Lambda_{QCD}^{n_{f}=2}$ in the range $[250, 350]$~MeV. We also tried other
options to evaluate $\alpha_{s}^{n_{f}=2}$~\cite{futuro}. 
In the numerical evaluation, we also used eq.~(\ref{eq:massa}) 
expanded to ${\cal O}(\alpha_{s}^{3})$. In this way, we obtain a
rough estimate of higher-order terms which is found less than $3\%$. 
Chiral extrapolations are also under control because the b--quark masses 
obtained from the $B_{d}$ and $B_{s}$ mesons are nicely compatible. Results from 
different smearing methods are also compatible, showing that the ground state
has been well isolated. Finally, the dependence on $k_{sea}$ is very small. This \
allows us to take for our best estimate of $\overline{m}_{b}(\overline{m}_{b})$
the value obtained at the lightest one,
$k_{sea}=0.1580$. Taking into account all errors, we find
\be\label{eq:ournumber}
 \overline{m}_{b}(\overline{m}_{b}) = 
 (4.26\pm 0.03 \pm 0.04 \pm 0.05)\;\;\mbox{\rm GeV}
\ee
where the first error is statistical, the second is systematic and the third is
an estimate of the effects of higher-order corrections in the matching.
 It is interesting to compare our result 
with our old quenched numbers~\cite{ours} at three different lattice spacings 
$a^{-1}=2.0(2), 2.9(3)$ and
$3.8(3)$ GeV, which have been reanalyzed in \cite{MS} to include the 
two--loop correction to the matching \cite{MS},
\be\label{eq:oldvalues}
\overline{m}_{b}(\overline{m}_{b}) = \{ 4.34(5), 4.29(7), 4.25(7) \}\;\;\mbox{\rm GeV}
\ee
where the error comes from the lattice uncertainties. In this case, an 
educated estimate of higher-order corrections is $0.10$ GeV. 
As can be seen, they are in good agreement with our unquenched result. Given 
the present uncertainties, in particular the small physical volume used 
to obtain the result for $a^{-1}=3.8$ GeV, we are not in the position to 
attempt the extrapolation of the results in~(\ref{eq:oldvalues}).  

\end{document}